\documentclass[journal]{IEEEtran}
\IEEEoverridecommandlockouts
\usepackage{makecell}
\usepackage{cite}
\usepackage{amsmath,amssymb,amsfonts}

\usepackage{algorithmic}
\usepackage{graphicx}
\usepackage{textcomp}

\newtheorem{remark}{Remark}
\usepackage{bbm}
\usepackage{bm}

\usepackage{bbm}
\usepackage{pifont}
\usepackage[table]{xcolor}

\usepackage{svg}

\makeatletter
\let\ps@IEEEtitlepagestyle\ps@empty
\let\ps@headings\ps@empty
\makeatother

\def\BibTeX{{\rm B\kern-.05em{\sc i\kern-.025em b}\kern-.08em
   T\kern-.1667em\lower.7ex\hbox{E}\kern-.125emX}}

\begin{document}
\setlength{\textfloatsep}{0.11cm}
\setlength{\abovedisplayskip}{0.1cm}
\setlength{\belowdisplayskip}{0.1cm}
\setlength{\baselineskip}{0.42cm}
\setlength{\abovecaptionskip}{1mm}

\title{{Constrained Capacity for Faster-than-Nyquist Signaling in Frequency-Selective Fading Channels}}
\author{Zichao~Zhang,~\IEEEmembership{Student Member,~IEEE,}
		Melda~Yuksel,~\IEEEmembership{Senior Member,~IEEE,}
  Gokhan M. Guvensen,
Halim~Yanikomeroglu,~\IEEEmembership{Fellow,~IEEE}
\thanks{This work was funded in part by a Discovery Grant awarded by the Natural Sciences and Engineering Research Council of Canada (NSERC).}
\thanks{Z. Zhang and H. Yanikomeroglu are with the Department of Systems and Computer Engineering at Carleton University, Ottawa, ON, K1S 5B6, Canada e-mail:	zichaozhang@cmail.carleton.ca, halim@sce.carleton.ca.}
\thanks{M. Yuksel and G. Guvensen are with the Department of Electrical and Electronics Engineering, Middle East Technical University, Ankara, 06800, Turkey, e-mail: {ymelda, guvensen}@metu.edu.tr.}
}

\maketitle
\thispagestyle{empty}
\enlargethispage{-0.38in}

\begin{abstract}
 %In this paper, we investigate the constrained capacity of discrete Fourier transform (DFT) precoded faster-than-Nyquist (FTN) signaling over frequency-selective channels with finite-alphabet inputs. By employing a cyclic prefix (CP) and cyclic suffix (CS), the intersymbol interference (ISI) from FTN and multipath channel is decomposed into parallel eigenchannels, whose gains are determined by together the folded spectrum of FTN and the channel frequency response. Based on this decomposition, we derive the constrained capacity. We further study the mismatched achievable information rate for the case where CP and CS are missing to show the effect of imperfect diagonalization and residual inter-eigenchannel interference under short packet lengths. %Numerical results show that, even under mismatched detection without CP and CS, DFT-precoded FTN with 16QAM achieves a rate improvement of approximately $X\%$ over the practical Nyquist RRC benchmark.
 In this paper, we investigate the constrained capacity of discrete Fourier transform (DFT)-precoded faster-than-Nyquist (FTN) signaling over frequency-selective channels with finite-alphabet inputs. With a cyclic prefix (CP) and cyclic suffix (CS), the FTN and multipath induced intersymbol interference (ISI) is decomposed into parallel eigenchannels, whose gains are jointly determined by the folded FTN spectrum and the channel frequency response. Based on this decomposition, we derive the constrained capacity for finite-alphabet constellations and formulate a mismatched decoding achievable information rate for DFT-precoded FTN signaling without CP/CS, quantifying the finite-block rate loss caused by imperfect diagonalization. We find that even under mismatched decoding, FTN significantly improves upon Nyquist transmission.
\end{abstract}


\section{Introduction}

The rapid growth of communication services is placing unprecedented pressure on radio spectrum resources. Early 6G vision documents identify demanding use cases such as immersive communication, global broadband, omnipresent Internet of Things (IoT), spatio-temporal services, critical services, and compute-artificial intelligence (AI) services \cite{whatshould6Gbe}. These emerging services require substantially higher traffic capacity, motivating transmission techniques with improved spectral efficiency.

Faster-than-Nyquist (FTN) signaling is a promising approach to improve spectral efficiency without increasing the occupied bandwidth. Since Mazo's seminal work \cite{mazo}, FTN has attracted considerable attention because it transmits symbols faster than the Nyquist no-intersymbol interference (ISI) rate. Let $T$ denote the Nyquist symbol interval. FTN uses a compressed interval $\delta T$, where $\delta\in(0,1]$ is the acceleration factor. Although this increases the symbol rate under the same bandwidth, it intentionally introduces ISI and therefore complicates detection.

To reduce receiver complexity, transmitter-side precoded FTN has been widely studied. Instead of directly transmitting information-bearing time-domain symbols, practical constellation symbols, such as quadrature phase-shift keying (QPSK) and quadrature amplitude modulation (QAM), are first loaded in the frequency domain and then precoded into the time domain. After reception, the signal is transformed back to the frequency domain, where the system can be interpreted as a set of parallel channels. This structure simplifies FTN detection and is attractive for practical implementation \cite{1_R1}. Precoding for FTN has also been studied in other aspects of FTN \cite{hongotfs,precodftnisac,impercsiftnprec,specftngtmhprecod}.

The capacity of FTN signaling has been studied through eigenvalue decomposition in \cite{property}, providing an important theoretical basis for precoded FTN transmission. Based on this perspective, singular value decomposition (SVD)-based FTN was proposed in \cite{svd}. In addition, \cite{timelocalization} proposed precoding based on the inverse square root of the FTN ISI matrix and showed that precoded FTN can be capacity-achieving under Gaussian signaling. Among practical precoders, the discrete Fourier transform (DFT) matrix is particularly attractive because it can diagonalize the FTN ISI matrix with the aid of cyclic prefix (CP) and cyclic suffix (CS) \cite{gray}.

Despite extensive studies on FTN capacity with Gaussian inputs, constrained-capacity analysis with finite-alphabet constellations remains limited. This gap is important because practical systems employ discrete modulations such as binary phase-shift keying (BPSK), QPSK, and QAM. Constrained capacity was classically studied by Ungerboeck for finite-alphabet channels \cite{ungerboeckconstrained} and later extended to FTN signaling in \cite{rusek}, where bounds were derived for finite-alphabet time-domain symbols. The asymptotic behavior of binary FTN signaling as $\delta\to0$ was further investigated in \cite{asympftn}. \enlargethispage{-0.38in}

In this paper, we investigate the constrained capacity of DFT-precoded FTN signaling with finite-alphabet inputs over frequency-selective (FS) fading channels. The FS channel captures multipath propagation with distinct tap gains and delays. With CP and CS, the combined FTN and multipath-induced ISI can be decomposed into parallel eigenchannels, which enables constrained-capacity evaluation for practical constellations. We further study the case without CP/CS, where finite-block DFT precoding cannot perfectly diagonalize the Toeplitz FTN ISI matrix. The resulting residual inter-eigenchannel interference leads to a mismatched decoding problem when the receiver still applies a simplified parallel-channel detector. Following the mismatched achievable information rate (AIR) framework in \cite{Gokhanmismatch}, we quantify the rate degradation caused by imperfect diagonalization. This work provides a practical information-theoretic framework for DFT-precoded FTN over FS fading channels, jointly accounting for finite-alphabet signaling, multipath propagation, finite-block effects, and mismatched detection.

\vspace{-0.15in}
\section{System Model}\label{sec:systemmodel}
\enlargethispage{-0.39in}

Assume that $N$ symbols are transmitted.
The independent and identically distributed (i.i.d.) data symbols are denoted by $\bm{s}=[s_0,s_1,\dots,s_{N-1}]^T$, where each symbol is drawn from a finite alphabet $\mathcal{S}$ of size $M$. Typical examples include QPSK with $M=4$ and 16QAM with $M=16$. Throughout this paper, we assume unit-energy constellation symbols; i.e., $\mathbb{E}[\|s_i\|^2]=1,i=0,\dots,N-1$,
and statistical independence among different symbols, namely $s_i \perp\!\!\!\perp s_j$ for $i\neq j$. Let $E_s$ denote the allocated symbol energy. Under uniform power allocation, the frequency-domain symbol vector is precoded by the DFT matrix before pulse shaping.

Let $\bm{D}$ denote the $N\times N$ normalized DFT matrix, whose entries are given by
\begin{equation}
    (\bm{D})_{k,n}=\frac{1}{\sqrt{N}}e^{-j\frac{2\pi}{N}kn},
    \qquad k,n=0,\dots,N-1.
\end{equation}
Let $\bm{a}=[a[0],\dots,a[N-1]]^T$  denote the precoded transmit vector, which is written as
\begin{equation}
    \bm{a}=\sqrt{E_s}\bm{D}\bm{s}.
\end{equation}
The precoded symbols $a[n]$ are modulated with the pulse-shaping filter $p(t)$ and transmitted at the accelerated symbol interval $\delta T$, where $T$ denotes the Nyquist symbol period and $\delta \in (0,1]$ is the acceleration factor. Throughout this paper, we assume that the pulse-shaping filter $p(t)$ is a root-raised cosine pulse with roll-off factor $\beta$. However, the analysis of this paper is not limited to root-raised cosine pulses and other pulse shapes are applicable as well. The transmitted signal $x(t)$ is therefore 
\begin{equation}
    x(t)=\sum_{n=0}^{N-1}  a[n]\, p\bigl(t-n\delta T\bigr).
\end{equation}

We assume a frequency-selective fading channel together with additive white complex Gaussian noise (AWGN) $n(t)$ at the receiver. The noise $n(t)$ has power spectral density $\sigma_0^2$. 
 The FS channel is modeled as a tapped-delay line with $J$ taps, where the $j$-th tap arrives at delay $j\delta T$ with tap gain $h_j$ \cite{freqselective}.
The signal at the receiver $r(t)$ will be 
\begin{equation}
r(t)=\sum_{n=0}^{N-1}\sum_{j=0}^{J-1} h_j a[n]\, p\bigl(t-(n+j)\delta T\bigr)+n(t).
\label{eq:fs_ct_signal}
\end{equation}
  At the receiver, a matched filter $p^*(-t)$ is employed to maximize the signal-to-noise ratio (SNR) at the sampling instants. Define the overall pulse shape $g(t)$ as 
    $g(t)=p(t)\star p^*(-t)$.
 Then the output of the receiver matched filter, $y(t)$, can be written as
\begin{align}
    y(t)=\sum_{n=0}^{N-1}\sum_{j=0}^{J-1} h_j a[n]\, g\bigl(t-(n+j)\delta T\bigr)+\eta(t), 
\end{align}
where $\eta(t)$ is  the filtered noise process. It is defined as
    $\eta(t)=n(t)\star p^*(-t)$.
By sampling at $m\delta T, m=0,\dots,N-1,$ then
\begin{equation}
y[m]=\sum_{n=0}^{N-1}\sum_{j=0}^{J-1} h_j a[n]\, g[m-n-j]+\eta[m],
\label{eq:fs_sampled_scalar}
\end{equation}
where $g[\ell]\triangleq g(\ell\delta T)$ and $\eta[m]$ is the sampled noise term and its covariance is given by 
$\mathbb{E}[\boldsymbol{\eta}\boldsymbol{\eta}^\dagger]=\sigma_0^2 \mathbf{G}$.
The matrix $\bm{G}$ is the ISI matrix with entries given by
\begin{equation}
    (\bm{G})_{k,\ell}=g((k-\ell)\delta T), \qquad k,\ell=0,\dots,N-1.
\end{equation}
Note that $\bm{G}$ is an $N\times N$ Hermitian Toeplitz matrix. %Since in FTN signaling, we have $g((i-j)\delta T)\neq0$, the $\bm{G}$ matrix is generally not a diagonal matrix, we can see that the noise is correlated. 
We define 
\begin{align}
    (\bm{G}^j)_{k,\ell}=g((k-\ell-j)\delta T),\qquad j=0,\dots,J-1. \label{eqn:Gjentrydef}
\end{align}
It is clear that matrix $\bm{G}^j$ is also a Toeplitz matrix and $\bm{G}^j|_{j=0}=\bm{G}$. 
Then, we can write the input-output relationship in \eqref{eq:fs_sampled_scalar} as 
\begin{align}
    \bm{y}=\sum_{j=0}^{J-1}h_j\bm{G}^j\bm{a}+\bm{\eta}=\bm{C}\bm{a}+\bm{\eta},\label{eqn:nocpcssysmodel}
\end{align}
where $\bm{y},\bm{\eta}\in\mathbb{C}^N$ and $\bm{G}^j\in\mathbb{C}^{N\times N}$,  
and the new Toeplitz matrix $\bm{C}$ and its entries $(\bm{C})_{k,\ell}$ are defined as 
\begin{eqnarray}
\bm{C}&=&\sum_{j=0}^{J-1}h_j\bm{G}^j, \label{eqn:Cdef} \\
    (\bm{C})_{k,\ell}&=&\sum_{j=0}^{J-1}h_jg((k-\ell-j)\delta T).\label{eqn:defC}
\end{eqnarray}
 
\enlargethispage{-0.39in}
We assume that the energy of pulse $g(t)$ is concentrated in interval $[-L\delta T, L\delta T]$, which means we assume that $g[\ell]\approx0, \forall|\ell|>L$. Thus, the matrix $\bm{G}$ is a banded Toeplitz matrix with only the middle $2L+1$ diagonals being non-zero. Similarly, the ISI matrices $\bm{G}^j$ corresponding to the delayed taps are also banded with length $2L+1$. According to \eqref{eqn:Gjentrydef}, we can see that if we set the main diagonal of an $N\times N$ matrix to be zeroth diagonal, and the upper right diagonal is $N$th, we can see that for $\bm{G}^j$ its $(-L-j)$th to $(L-j)$th diagonals are not zero, and it is a shifted version of $\bm{G}$. By inspecting \eqref{eqn:Cdef}, we can see that $\bm{C}$ is also banded Toeplitz matrix. Although it is not Hermitian, its $(-L-J+1)$th to $L$th diagonals are non-zero. Therefore, when choosing the length for CP and CS, we need the length of CP to be $L+J$ and the length of CS to be $L$ to make $\bm{C}$ cyclic. After CP and CS, we denote the equivalent cyclic matrix by $\bm{C}_c$.

\begin{figure}
    \centering
    \includegraphics[width=0.75\linewidth]{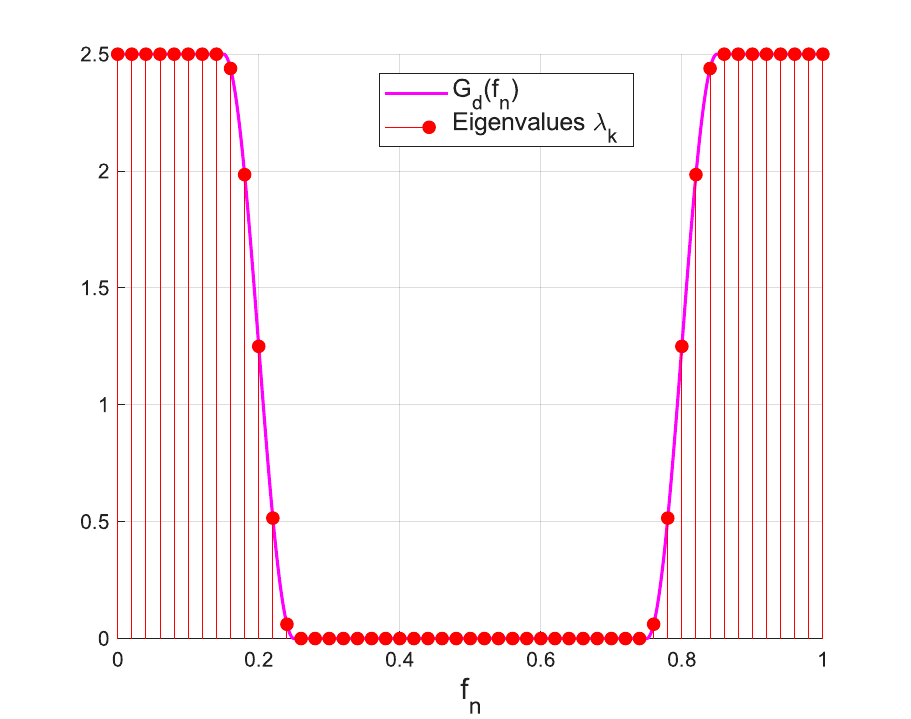}
    \caption{Folded spectrum $G_d(f_n)$ over one normalized frequency period of $[0,1)$ and the corresponding eigenvalues $\lambda_i$ of the circulant FTN ISI matrix $\bm{G}_c$, where $\delta=0.4$ and $\beta=0.25$.}
    \label{fig:eigenvalfolspect}
\end{figure}

To obtain a parallel channel representation, we multiply the output $\bm{y}$ with the IDFT matrix $\bm{D}^\dagger$, namely, 
\begin{align}
    \tilde{\bm{y}}=\bm{D}^\dagger\bm{y}=\bm{D}^\dagger\bm{C}_c\bm{D}\bm{s}+\bm{D}^\dagger\bm{\eta}.
\end{align}
According to \cite{gray}, the DFT matrix is the eigenmatrix for circulant matrices, therefore $\bm{C}_c$ can be diagonalized as
\begin{align}
    \tilde{\bm{\Lambda}}=\bm{D}^\dagger\bm{C}_c\bm{D}=\text{diag}\{\tilde{\lambda}_0, \dots, \tilde{\lambda}_{N-1}\}.
\end{align}
Denote the $\ell$th diagonal value of $\bm{C}_c$ as $c[\ell]$, which is also the discrete-time response
\begin{equation}
    c[\ell]\triangleq \sum_{j=0}^{J-1} h_j\, g[\ell-j].
\end{equation}
Since $g[n]=0$ for $|n|>L$, the support of $c[\ell]$ is
    $\ell\in[-L,\;L+J-1]$.
The first row of $\bm{C}_c$ is
    $\bigl[
    c[0],\; c[-1],\; \dots,\; c[-L],\;\allowdisplaybreaks 0,\; \dots,\; 0,\; c[L+J-1],\; c[L+J-2],\; \dots,\; c[1]
    \bigr]$.
The $k$th diagonal entry of $\tilde{\bm{\Lambda}}$ is generated by the DFT of the first row of $\bm{C}_c$, namely
\begin{equation}
    \tilde{\lambda}_k
    =\sum_{\ell=-L}^{L+J-1} c[\ell]\,e^{-j\frac{2\pi}{N}k\ell},
    \qquad k=0,1,\dots,N-1.
\end{equation}
Substituting $c[\ell]=\sum_{j=0}^{J-1} h_j g[\ell-j]$, we obtain
\begin{align}
    \tilde{\lambda}_k
    &=\sum_{\ell=-L}^{L+J-1}\sum_{j=0}^{J-1}
    h_j\,g[\ell-j]\,
    e^{-j\frac{2\pi}{N}k\ell}.
\end{align}
Let $n=\ell-j$, and since $g[n]=0$ for $|n|>L$, we get
\begin{align}
    \tilde{\lambda}_k
    &=\sum_{j=0}^{J-1}\sum_{n=-L}^{L}
    h_j\,g[n]\,
    e^{-j\frac{2\pi}{N}k(n+j)}.
\end{align}
Hence,
\begin{equation}
    \tilde{\lambda}_k
    =
    \left(\sum_{n=-L}^{L} g[n] e^{-j\frac{2\pi}{N}kn}\right)
    \left(\sum_{j=0}^{J-1} h_j e^{-j\frac{2\pi}{N}kj}\right). \label{eqn:Cclambdecomposed}
\end{equation}
We denote the FTN spectrum coefficients as 
\begin{equation}
    \lambda_k \triangleq \sum_{n=-L}^{L} g[n] e^{-j\frac{2\pi}{N}kn}, \label{eqn:ftnfolspect}
\end{equation}
and we denote the FS spectrum coefficients  as 
\begin{equation}
    H_k \triangleq \sum_{j=0}^{J-1} h_j e^{-j\frac{2\pi}{N}kj}.
\end{equation}
Then the $k$th diagonal entry of $\tilde{\bm{\Lambda}}$ can be written as
\enlargethispage{-0.38in}
\begin{equation}
    \tilde{\lambda}_k=\lambda_k H_k,
\end{equation}
and the diagonal matrix $\tilde{\bm{\Lambda}}$ is
\begin{align}
    \tilde{\bm{\Lambda}}
    &=
    \mathrm{diag}\{\lambda_0 H_0,\lambda_1 H_1,\dots,\lambda_{N-1}H_{N-1}\}=\bm{\Lambda}\bm{H}.
\end{align}
By observing \eqref{eqn:ftnfolspect}, the coefficients $\lambda_k$ are samples from the folded spectrum $G_d(f_n)$ and are defined as 
\begin{align}
    G_d(f_n)
    &=\sum_{n=-L}^{L}g[n]e^{j2\pi f_nn} \label{eqn:deffoldspectsum}\\
    &=\frac{1}{\delta T}\sum_{k=-\infty}^{+\infty}G\!\left(\frac{f_n-k}{\delta T}\right),
    \quad
    f_n\in\left[-\frac{1}{2},\frac{1}{2}\right], \label{eqn:deffoldspectGf}
\end{align}
where $G(f)$ is the continuous-time Fourier transform (CTFT) of $g(t)$. From  \eqref{eqn:deffoldspectsum}, we can see that the folded spectrum $G_d(f_n)$ is a periodic spectrum with period 1. Moreover, \eqref{eqn:deffoldspectGf} suggests that the folded spectrum can be obtained by duplicating and shifting the CTFT $G(f)$ by $k/\delta T$, then scaling the frequency by $f_n=f/\delta T$. From \eqref{eqn:ftnfolspect} and \eqref{eqn:deffoldspectsum} we can see that eigenvalues $\tilde{\lambda}_k$ are simply    $\tilde{\lambda}_k=G_d(f_n)|_{f_n=\frac{k}{N}}, \quad k=0, \dots, N-1$,
and are equal to the samples of the folded spectrum in one period. 
We plot the folded spectrum for one period in Fig.~\ref{fig:eigenvalfolspect} together with the eigenvalues $\tilde{\lambda}_k$. 
As we can see, the overall eigenchannel coefficients are composed of two separate parts, the coefficients from FTN spectrum and the coefficients from the discrete channel frequency response of the FS taps. 
We can also see from Fig.~\ref{fig:eigenvalfolspect} that some of the samples are zero, and the fraction of nonzero eigenvalues is approximately determined by the support of $G_d(f_n)$ over one period \cite{property}.

On the other hand, after receiver processing, the equivalent noise becomes 
\begin{align}
    \bm{\omega}=\bm{D}^\dagger\bm{\eta},
\end{align}
and its covariance matrix is given as 
\begin{align}
    \mathbb{E}\left[\bm{\omega}\bm{\omega}^\dagger\right]=\mathbb{E}\left[(\bm{D}^\dagger\bm{\eta})(\bm{D}^\dagger\bm{\eta})^\dagger\right]=\sigma_0^2\bm{D}^\dagger\bm{G}\bm{D}. \label{eqn:covmatofomega}
\end{align}
Since $\bm{G}$ is not cyclic, the product $\bm{D}^\dagger\bm{G}\bm{D}$ is generally not diagonal, and the noise term $\bm{\omega}$ is still correlated. However, this effect will vanish when $N$ is sufficiently large. For sufficiently large $N$, the DFT matrix $\bm{D}$ is the eigenvector matrix for $\bm{G}$ and  $\bm{\Lambda}$ is the covariance matrix for $\bm{\omega}$, and $\bm{\omega}$ is uncorrelated.

\subsubsection{Power constraint}
Assume that the physical transmit-power limit is $P_{TX}$. For sufficiently large $N$, the overhead introduced by the CP and CS is negligible. Then, the average transmit power constraint can be written as
\begin{align}
    \mathbb{E}\left[\frac{1}{N\delta T}\int_{-\infty}^{\infty}|x(t)|^2dt\right] =\frac{1}{N\delta T}\mathrm{tr}(\bm{G}\bm{\Sigma}_a)
    \leq P_{TX},
    \label{eqn:powconst}
\end{align}
where $
    \bm{\Sigma}_a=\mathbb{E}[\bm{a}\bm{a}^\dagger]=E\bm{D}\mathbb{E}[\bm{s}\bm{s}^\dagger]\bm{D}^\dagger$.
It follows that
\begin{align}
    \frac{1}{N\delta T}\mathrm{tr}(\bm{G}\bm{\Sigma}_a)
    &=\frac{E_s}{N\delta T}
    \mathrm{tr}\!\left(
    \bm{G}\bm{D}\mathbb{E}[\bm{s}\bm{s}^\dagger]\bm{D}^\dagger
    \right)\notag\\
    &=\frac{E_s}{N\delta T}
    \mathrm{tr}\!\left(
    \bm{D}^\dagger\bm{G}\bm{D}\mathbb{E}[\bm{s}\bm{s}^\dagger]
    \right)\notag\\
    &\overset{(a)}{=}
    \frac{E_s}{N\delta T}\mathrm{tr}(\bm{\Lambda}).
    \label{eqn:powconstraint}
\end{align} \enlargethispage{-0.38in}
where (a) follows from the fact that for large enough $N$ \cite{gray}, we have the relationship 
    $\bm{D}^\dagger\bm{G}\bm{D}\approx\bm{D}^\dagger\bm{G}_c\bm{D}=\bm{\Lambda}$.
Now that the FS FTN channel is decomposed into parallel channels by the DFT matrix, we can write the $k$-th equivalent scalar subchannel as
\begin{equation}
\widetilde{y}_k=\sqrt{E_s}\,\lambda_k H_k s_k+\omega_k.
\label{eq:fs_branch_model}
\end{equation}
% The effective SNR of the $k$-th subchannel is 
% \begin{equation}
% \gamma_k^{\mathrm{FS}}
% =
% \frac{E_s|\lambda_k H_k|^2}{\sigma_0^2\lambda_k}
% =
% \frac{E_s\lambda_k |H_k|^2}{\sigma_0^2}.
% \label{eq:fs_branch_snr}
% \end{equation}

\section{The Constrained capacity derivation}

Given the input-output relationship in  \eqref{eq:fs_branch_model}, we can see that the symbol of each eigenchannel is scaled by $\lambda_kH_k$. 
Let $N_r$ denote the number of nonzero eigenvalues. As shown in Fig.~\ref{fig:eigenvalfolspect}, when $\delta(1+\beta)<1$, some  eigenvalues are zero. For sufficiently large $N$, the fraction of nonzero eigenvalues can be approximated as
\begin{equation}
    \frac{N_r}{N}\approx \min\bigl(1,\delta(1+\beta)\bigr).
    \label{eqn:NandNred}
\end{equation}
When $\lambda_k=0$, the $k$th eigenchannel is inactive since the symbol $s_k$ is scaled by zero.
Hence, when $\delta(1+\beta)<1$, only $N_r$ eigenchannels are active, while the remaining $N-N_r$ eigenchannels are effectively nulled by zero eigenvalues. In other words, only the active eigenchannels can carry useful information. Therefore, we modulate symbols only on the active eigenchannels, whose index set is
\begin{align}
    \mathcal{D}=\left\{0, \dots, \frac{N_r}{2},  N-\frac{N_r}{2}, \dots, N-1\right\}. \label{eqn:symmodulateindex}
\end{align}

Following the approach in \cite{ungerboeckconstrained}, we obtain the constrained capacity for the $k$th eigenchannel as in \eqref{eq:fs_branch_cc}, shown on top of the next page, where $\mathbb{E}_{\omega_k}[\cdot]$ means the expectation is taken over $\omega_k$ and $s_k[m]$ means the $m$th constellation point of symbol $s_k$. 

The constrained capacity of DFT-precoded SISO FTN for an FS channel is then obtained by summing the information carried over all active eigenchannels and averaging it over the transmitted time-domain symbols, namely,
\begin{figure*}
\begin{equation}
C_k^{\mathrm{FS}}(\delta)
=
\log_2 M
-
\frac{1}{M}\sum_{m=1}^{M}
\mathbb{E}_{\omega_k}\!\left[
\log_2
\sum_{\ell=1}^{M}
\exp\!\left(
-\frac{
\left|\omega_k+\sqrt{E_s}\lambda_k H_k\bigl(s_k[m]-s_k[\ell]\bigr)\right|^2
-
|\omega_k|^2
}{
\sigma_0^2\lambda_k
}
\right)
\right].
\label{eq:fs_branch_cc}
\end{equation}
\vspace{-0.25in}
\end{figure*}
\begin{equation}
    C^{FS}(\delta)=\frac{1}{N}\sum_{d\in\mathcal{D}} C^{FS}_d(\delta),
    \label{eqn:persymcap}
\end{equation} in bits per transmitted symbol. The key advantage of FTN signaling; however, lies in its reduced transmission interval in time. Specifically, transmitting $N$ time-domain symbols requires a total duration of only $N\delta T$, while only $N_r$ frequency-domain symbols effectively carry useful information. As $\delta$ decreases, \eqref{eqn:NandNred} shows that the ratio $N_r/N$ becomes smaller, so the information carried per time-domain symbol decreases. Nevertheless, the corresponding transmission duration is also reduced, which improves the overall spectral efficiency. Therefore, the constrained capacity in bits/s/Hz is given by
\begin{equation}
    \tilde{C}^{FS}(\delta)=\frac{1}{\delta(1+\beta)}C^{FS}(\delta).
    \label{eqn:capbpsphz}
\end{equation}
Substituting \eqref{eqn:NandNred} into \eqref{eqn:capbpsphz}, we obtain
\begin{equation}
    \tilde{C}^{FS}(\delta)
    =
    \frac{1}{N_r}
    \sum_{d\in\mathcal{D}} C^{FS}_d(\delta),
    \label{eqn:aveovereigenchnl}
\end{equation}
which shows that the spectral efficiency in bits/s/Hz is equal to the average constrained capacity over all active eigenchannels.
\begin{remark}
    For FTN signaling under transmit power $P_{\rm TX}$, the symbol energy is $E_s=P_{\rm TX}\delta T$. Therefore, we define
$\mathsf{SNR_{tx}}=P_{\rm TX}/\sigma_0^2$ and
$\mathsf{SNR_{rx}}=(E_s/T)/\sigma_0^2=\delta P_{\rm TX}/\sigma_0^2$,
which coincide only for Nyquist signaling with $\delta=1$, \cite{capiapr}.
\end{remark}
\enlargethispage{-0.38in}

\section{Mismatched AIR for FTN Signaling}

Although CP and CS enable an exact decomposition of the FS-FTN channel, their required lengths depend on both the FTN pulse memory and the multipath delay spread. For low-latency transmission, this guard overhead may be non-negligible, and practical systems may therefore employ shortened guard intervals or omit them. In this case, the effective channel matrix remains Toeplitz rather than circulant, so DFT-based detection leads to residual inter-eigenchannel interference and a mismatched decoding problem.
%\eqref{eqn:nocpcssysmodel}-\eqref{eqn:defC}
Applying $\bm{D}^\dagger$ to the received vector yields 
\begin{align}
    \tilde{\bm y}
    &=\bm D^\dagger (\bm C \bm a+\bm\eta) \notag\\
    &=\bm D^\dagger \bm C \sqrt{E_s}\bm D \bm s+\bm D^\dagger \bm\eta \notag
\end{align}
\begin{align}
    &=\sqrt{E_s}\,\bm\Gamma\bm s+\bm \omega,
    \label{eq:fs_freq_model}
\end{align}
where
    $\bm\Gamma\triangleq \bm D^\dagger \bm C \bm D$, and 
    $\bm\omega\triangleq \bm D^\dagger \bm\eta$.
% The $(k,\ell)$th entry of $\bm\Gamma$ is
% \begin{align}
%     (\bm\Gamma)_{k,\ell}
%     &=
%     \sum_{m=0}^{N-1}\sum_{n=0}^{N-1}
%     [\bm D^\dagger]_{k,m} C_{m,n} [\bm D]_{n,\ell} \notag\\
%     &=
%     \frac{1}{N}\sum_{m=0}^{N-1}\sum_{n=0}^{N-1}
%     c[m-n]e^{-j\frac{2\pi}{N}km}e^{j\frac{2\pi}{N}\ell n}.
%     \label{eq:fs_Gamma_entry_1}
% \end{align}
% Let
% \begin{equation}
%     r=m-n.
%     \label{eq:fs_r_def}
% \end{equation}
% Then $m=n+r$, and for a fixed $r$, the valid values of $n$ satisfy
% \begin{equation}
%     \mathcal I_r=\left\{
%     n:\max(0,-r)\le n\le \min(N-1,N-1-r)
%     \right\}.
%     \label{eq:fs_Ir_def}
% \end{equation}
% Hence,
% \begin{align}
%     (\bm\Gamma)_{k,\ell}
%     &=
%     \frac{1}{N}\sum_{r=-(N-1)}^{N-1}
%     c[r]e^{-j\frac{2\pi}{N}kr}
%     \sum_{n\in\mathcal I_r}
%     e^{-j\frac{2\pi}{N}(k-\ell)n}.
%     \label{eq:fs_Gamma_entry_2}
% \end{align}
For finite block length, $\bm\Gamma$ is generally non-diagonal, and its off-diagonal entries quantify the residual inter-eigenchannel interference.

As calculated in \eqref{eqn:covmatofomega} the covariance matrix for $\bm{\omega}$ is   
    $\sigma_0^2 \bm D^\dagger \bm G \bm D
    \triangleq
    \sigma_0^2\bm K$,
where $\bm K\triangleq \bm D^\dagger \bm G \bm D$. Now, $N$ is not large enough and the DFT matrix is no longer the eigenvector matrix for $\bm{G}$ anymore. 
% Its $(k,\ell)$th entry is
% \begin{align}
%     K_{k,\ell}
%     &=
%     \frac{1}{N}\sum_{m=0}^{N-1}\sum_{n=0}^{N-1}
%     g[m-n]e^{-j\frac{2\pi}{N}km}e^{j\frac{2\pi}{N}\ell n}.
%     \label{eq:fs_K_entry_1}
% \end{align}
% Using the same change of variable $r=m-n$, we obtain
% \begin{align}
%     K_{k,\ell}
%     &=
%     \frac{1}{N}\sum_{r=-(N-1)}^{N-1}
%     g[r]e^{-j\frac{2\pi}{N}kr}
%     \sum_{n\in\mathcal I_r}
%     e^{-j\frac{2\pi}{N}(k-\ell)n}.
%     \label{eq:fs_K_entry_2}
% \end{align}
Thus, $\bm{K}$ is not a diagonal matrix, and the output noise is correlated for finite $N$.

 Define the diagonal selection matrix $\bm\Pi_{\mathcal D}$ whose diagonal entries are one on $\mathcal D$ and zero otherwise. Then the transmit covariance matrix is 
\begin{equation}
    \bm\Sigma_a
    =
    E_s\bm D \bm\Pi_{\mathcal D}\bm D^\dagger.
    \label{eq:fs_Sigma_a}
\end{equation}\enlargethispage{-0.38in}
Since the physical transmit-power constraint applies to the transmitter, the average transmit power is given by
substituting \eqref{eq:fs_Sigma_a} in \eqref{eqn:powconst}, and we have
\begin{align}
    \frac{1}{N\delta T}\operatorname{tr}(\bm G \bm\Sigma_a)
    &=
    \frac{E_s}{N\delta T}\operatorname{tr}(\bm D^\dagger \bm G \bm D \bm\Pi_{\mathcal D}) \notag\\
    &=
    \frac{E_s}{N\delta T}\sum_{d\in\mathcal D}(\bm{K})_{d,d} \label{eq:fs_power_constraint_2}\\
    E_s&=\frac{NP_{TX}\delta T}{\sum_{d\in\mathcal D}(\bm{K})_{d,d}}.
    \label{eq:fs_symbol_energy}
\end{align}
From \eqref{eq:fs_freq_model}, the $d$th active subchannel can be written as
\begin{equation}
    \tilde y_d
    =
    \sqrt{E_s}\,(\bm{\Gamma})_{d,d}s_d
    +
    \sqrt{E_s}\sum_{\substack{\ell\neq d\\ \ell\in\mathcal D}}
    (\bm{\Gamma})_{d,\ell}s_\ell
    +
    \omega_d.
    \label{eq:fs_branch_model2}
\end{equation}
We also define the residual interference term
\begin{equation}
    \xi_d
    \triangleq
    \sqrt{E_s}\sum_{\substack{\ell\neq d\\ \ell\in\mathcal D}}
    (\bm{\Gamma})_{d,\ell}s_\ell.
    \label{eq:fs_xi_def}
\end{equation}
% \begin{align}
%     \mathbb E[|\xi_d|^2]
%     &=
%     E_s\,
%     \mathbb E\!\left[
%     \left|
%     \sum_{\substack{\ell\neq d\\ \ell\in\mathcal D}}
%     (\bm{\Gamma})_{d,\ell}s_\ell
%     \right|^2
%     \right] \notag\\
%     &=
%     E_s\,
%     \mathbb E\!\left[
%     \sum_{\substack{\ell\neq d\\ \ell\in\mathcal D}}
%     \sum_{\substack{q\neq d\\ q\in\mathcal D}}
%     (\bm{\Gamma})_{d,\ell}(\bm{\Gamma})_{d,q}^*
%     s_\ell s_q^*
%     \right].
% \end{align}
Since the symbols are independent, zero mean, and have unit energy,
 all cross terms vanish, and the variance of residual interference term becomes
\begin{equation}
    \mathbb E[|\xi_d|^2]
    =
    E_s\sum_{\substack{\ell\neq d\\ \ell\in\mathcal D}}
    |(\bm{\Gamma})_{d,\ell}|^2.
    \label{eq:fs_interference_var}
\end{equation}
Similarly, the noise variance of the $d$th eigenchannel is
\begin{equation}
    \mathbb E[|\omega_d|^2]
    =
    \sigma_0^2(\bm{K})_{d,d}.
    \label{eq:fs_noise_branch_var}
\end{equation}
The mismatched decoder assumes the eigenchannels are uncorrelated and takes the inter-eigenchannel interference as noise. Thus, the approximated eigenchannel model is given as
\begin{equation}
    \tilde y_d
    \approx
    \sqrt{E_s}\,(\bm{\Gamma})_{d,d}s_d+\tilde\omega_d,
    \label{eq:fs_aux_model}
\end{equation}
where
    $\tilde\omega_d\sim \mathcal{CN}(0,\nu_{d}^2)$
with the variance 
\begin{equation}
    \nu_{d}^2
    =
    \sigma_0^2(\bm{K})_{d,d}
    +
    E_s\sum_{\substack{\ell\neq d\\ \ell\in\mathcal D}}
    |(\bm{\Gamma})_{d,\ell}|^2.
    \label{eq:fs_aux_var}
\end{equation}

\begin{figure}
    \centering
     \includegraphics[width=0.75\linewidth]{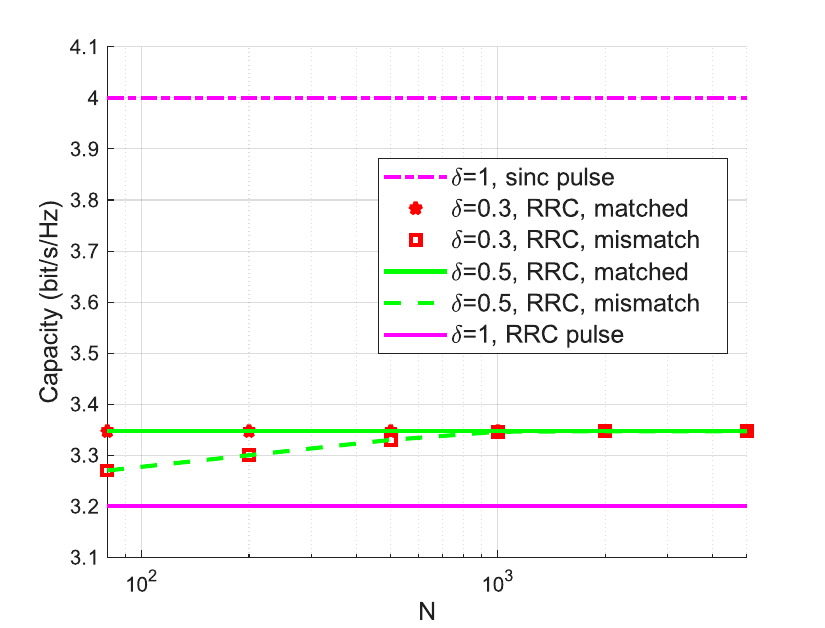}
    \caption{Constrained capacity vs mismatched AIR for 16QAM for different block length $N$, where $\mathsf{SNR_{tx}}=P_{TX}/\sigma_0^2$ is fixed. For two different $\delta$ values we compare with the Nyquist cases for both sinc and RRC pulses.}
    \label{fig:misvamatvsNtxsnr}
\end{figure}

We now formulate the mismatched AIR following the approach in \cite{Gokhanmismatch}. The mismatched receiver detects each active 
eigenchannel independently according to the scalar
model in \eqref{eq:fs_aux_model}. For a candidate symbol $a\in\mathcal S$, the receiver has the likelihood
\begin{equation}
    q_d(\tilde y_d|a)
    =
    \frac{1}{\pi \nu_d^2}
    \exp\left(
    -\frac{|\tilde y_d-\sqrt{E_s}(\bm{\Gamma})_{d,d}a|^2}{\nu_d^2}
    \right).
    \label{eq:fs_aux_likelihood}
\end{equation}
For uniformly distributed finite-alphabet inputs, if $\bar{a}$ is transmitted in the $d$th subchannel, the
mismatched information density \cite{Gokhanmismatch} of the $d$th active
eigenchannel is
\begin{equation}
    \iota_d(s_d,\tilde y_d)
    =
    \log_2 M
    -
    \log_2
    \left(
    \frac{\sum\limits_{a\in\mathcal S} q_d(\tilde y_d|s_d=a)}
    {q_d(\tilde y_d|s_d=\bar{a})}
    \right).
    \label{eq:mm_branch_info_density}
\end{equation}
Therefore, the mismatched AIR in bits per transmitted
time-domain symbol is \enlargethispage{-0.38in}
\begin{equation}
    I_{\mathrm{mm}}^{\mathrm{FS}}(\delta)
    =
    \mathbb{E}_{\tilde{\bm{y}}}
    \left[
    \frac{1}{N}
    \sum_{d\in\mathcal{D}}
    \iota_d(s_d,\tilde{y}_d)
    \right],
\end{equation}
where the expectation is taken with respect to the true
DFT-precoded FS-FTN distribution
\begin{equation}
    \tilde{\bm{y}} = \sqrt{E_s}\bm{\Gamma}\bm{s}+\bm{\omega},
    \qquad 
    \bm{\omega}\sim \mathcal{CN}(\bm{0},\sigma_0^2\bm{K}), \label{eqn:truefsmodel}
\end{equation}
whereas the likelihood inside $\iota_d(s_d,\tilde{y}_d)$ is
computed according to the mismatched scalar metric
$q_d(\tilde{y}_d|a)$. The corresponding mismatched AIR in bits/s/Hz is then given as 
\begin{equation}
    I_{\mathrm{mm}}^{\prime \mathrm{FS}}(\delta)
    =
    \frac{1}{\delta(1+\beta)}
    I_{\mathrm{mm}}^{\mathrm{FS}}(\delta). \label{eqn:mismatbpsphzorig}
\end{equation}
When $\delta(1+\beta)<1$ or $N_r/N\approx \delta(1+\beta)$,
\eqref{eqn:mismatbpsphzorig} can be approximated as
\begin{equation}
    I_{\mathrm{mm}}^{\prime \mathrm{FS}}(\delta)
    \approx
    \mathbb{E}
    \left[
    \frac{1}{N_r}
    \sum_{d\in\mathcal{D}}
    \iota_d(s_d,\tilde{y}_d)
    \right]. \label{eqn:misairbpsphz}
\end{equation}
Since the expectation is taken under the true correlated FS-FTN model given in \eqref{eqn:truefsmodel}, the mismatched AIR in \eqref{eqn:misairbpsphz} can be evaluated
numerically by Monte Carlo simulation. 

The AIR should be interpreted as a coded achievable rate rather than an uncoded detection metric. Specifically, for a sufficiently long and well-designed channel code using the same mismatched decoding metric $q_d(\tilde{y}_d|a)$, reliable transmission is possible for code rates below $I_{\rm mm}^{\rm FS}(\delta)$. Therefore, the gap between the constrained capacity and the mismatched AIR reflects the rate loss caused by imperfect diagonalization and residual inter-eigenchannel interference after the benefit of good channel coding is already accounted for. As $N$ increases, the Toeplitz FS-FTN channel matrix $\bm{C}$ becomes
asymptotically circulant, making the residual off-diagonal interference
after DFT processing negligible. Therefore, the mismatched AIR
approaches the constrained capacity in \eqref{eqn:aveovereigenchnl}.\enlargethispage{-0.38in}

\begin{figure}
    \centering
    \includegraphics[width=0.75\linewidth]{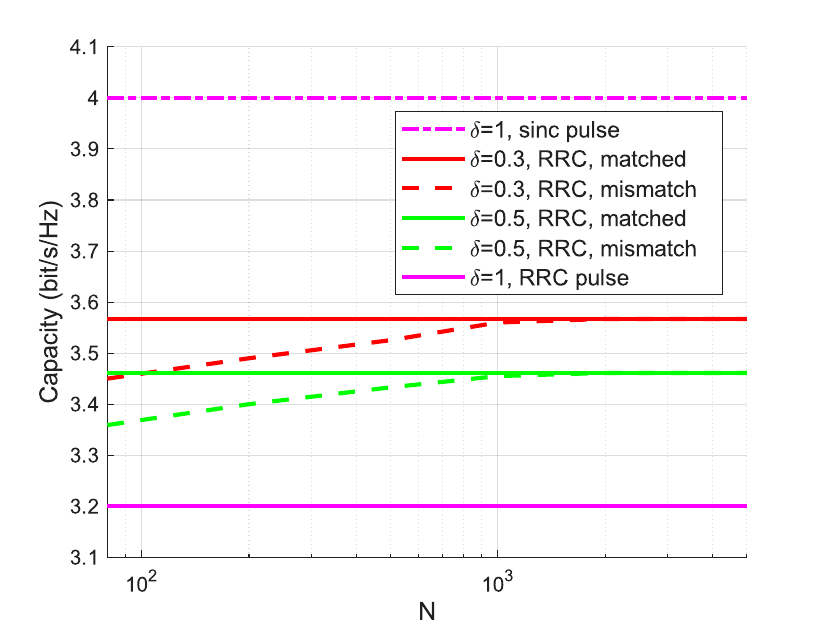}
    \caption{Constrained capacity vs mismatched AIR for 16QAM for different block length $N$, where $\mathsf{SNR_{rx}}=\delta P_{TX}/\sigma_0^2$ is fixed. For two different $\delta$ values we compare with the Nyquist cases for both sinc and RRC pulses.}
    \label{fig:misvsmatvsNrxsnr}
\end{figure}

\section{Simulation Results}

In this section, we evaluate the constrained capacity and mismatched AIR. Unless otherwise specified, we use an FS fading channel with $J=5$ taps and normalized average energy, where $h_j\sim\mathcal{CN}(0,1/J)$ for $j=0,\ldots,J-1$. The transmit pulse is an RRC pulse with $\beta=0.25$ and $T=1$. The Nyquist sinc pulse result is used as an ideal reference, while Nyquist RRC transmission is used as the practical benchmark.

We first compare the constrained capacity in \eqref{eqn:aveovereigenchnl} and the mismatched AIR in \eqref{eqn:misairbpsphz} versus the block length $N$. Fig.~\ref{fig:misvamatvsNtxsnr} shows the fixed $\mathsf{SNR_{tx}}$ results for 16QAM over the FS channel. The matched FTN curves outperform the Nyquist RRC benchmark because FTN uses the RRC excess bandwidth more efficiently. For small $N$, the mismatched AIR is below the constrained capacity since finite-block DFT precoding leaves residual inter-eigenchannel interference. As $N$ increases, this interference decreases and the mismatched AIR approaches the constrained capacity. The curves for $\delta=0.3$ and $\delta=0.5$ overlap because both satisfy $\delta(1+\beta)<1$ for $\beta=0.25$ and lie in the folded-spectrum saturation region, where the reduction $N_r/N\approx\delta(1+\beta)$ is compensated by the normalization factor $1/[\delta(1+\beta)]$~\cite{zhang2022faster,capiapr}. 

Fig.~\ref{fig:misvsmatvsNrxsnr} shows the corresponding fixed $\mathsf{SNR_{rx}}$ results. Again, FTN outperforms the Nyquist RRC benchmark by exploiting the RRC excess bandwidth. Unlike the fixed $\mathsf{SNR_{tx}}$ case, the curves for $\delta=0.3$ and $\delta=0.5$ do not overlap because fixed $\mathsf{SNR_{rx}}=E_s/(T\sigma_0^2)$ keeps the received symbol energy fixed, while $E_s=P_{ TX}\delta T$. Thus, a smaller $\delta$ requires a larger $P_{ TX}$ and gives the $\delta=0.3$ case a higher rate. For small $N$, the mismatched AIR remains below the constrained capacity due to residual off-diagonal interference, but it approaches the constrained capacity as $N$ increases. \enlargethispage{-0.38in}

\begin{figure}
    \centering
    \includegraphics[width=0.75\linewidth]{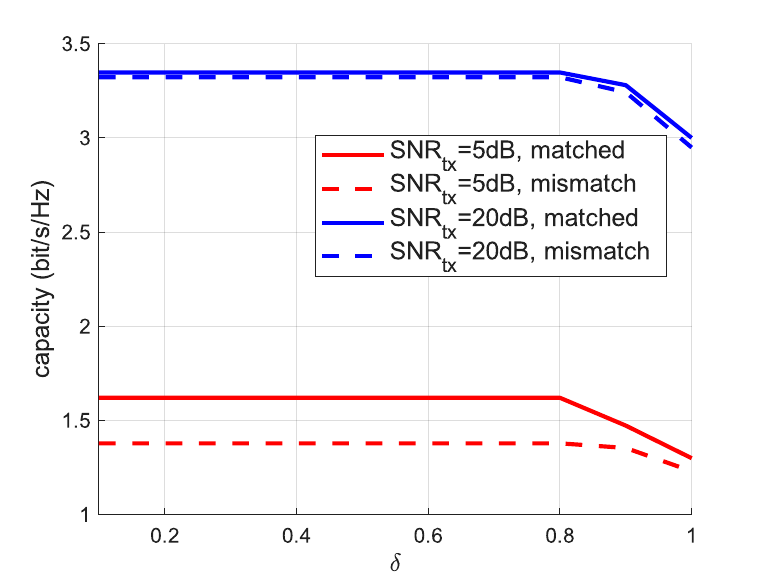}
    \caption{Constrained capacity vs mismatched AIR for 16QAM for different acceleration factor $\delta$, where $\mathsf{SNR_{tx}}=P_{TX}/\sigma_0^2$ is fixed.}
    \label{fig:captautx}
    \vspace{-0.15in}
\end{figure}
\begin{figure}
    \centering
    \includegraphics[width=0.75\linewidth]{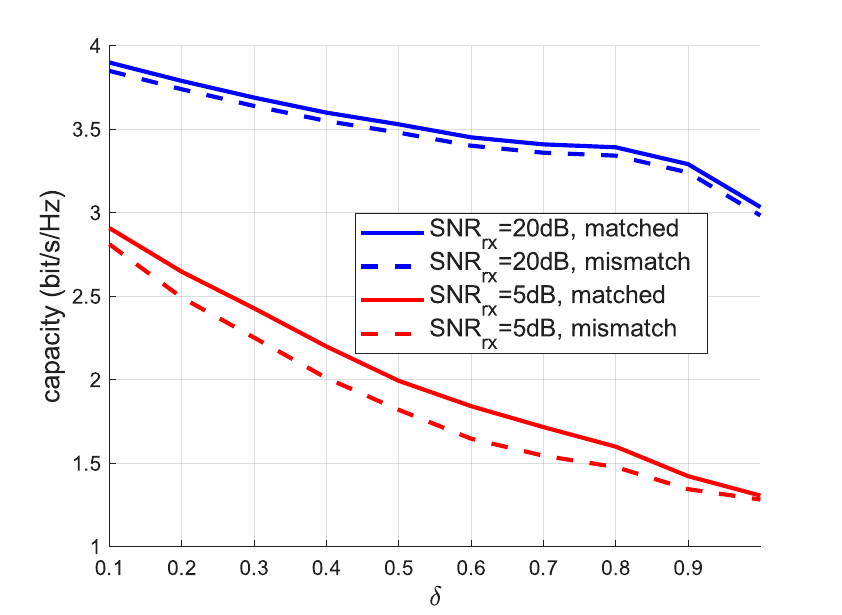}
    \caption{Constrained capacity vs mismatched AIR for 16QAM for different acceleration factor $\delta$, where $\mathsf{SNR_{rx}}=\delta P_{TX}/\sigma_0^2$ is fixed. }
    \label{fig:captaurx}
\end{figure}

Figs.~\ref{fig:captautx} and \ref{fig:captaurx} compare the constrained
capacity and mismatched AIR versus $\delta$ under fixed
$\mathsf{SNR_{tx}}$ and fixed $\mathsf{SNR_{rx}}$, respectively.
Under fixed $\mathsf{SNR_{tx}}$, the rate saturates as $\delta$
decreases because the finite alphabet and the number of active
eigenchannels limit the achievable spectral efficiency. Under fixed
$\mathsf{SNR_{rx}}$, the symbol energy is fixed, so decreasing
$\delta$ corresponds to a larger transmit power and the rate can
continue to increase after the active-eigenchannel threshold is reached.
The mismatch loss is more pronounced at low SNR, where noise and
residual inter-eigenchannel interference have a stronger impact on the
decoding metric.  

\vspace{-0.2cm}
\section{Conclusion}\label{sec:conclusion}
%In this paper, we investigated the constrained capacity of DFT-precoded FTN signaling over FS fading channels with finite-alphabet inputs. By employing CP and CS, the combined FTN ISI and multipath channel was transformed into parallel eigenchannels, which enabled the constrained-capacity analysis under practical constellations. We further considered the case without CP and CS and formulated the mismatched AIR caused by finite-block imperfect diagonalization and residual inter-eigenchannel interference. Simulation results showed that the mismatched AIR suffers from a performance loss for short block lengths, but gradually approaches the constrained capacity as the block length increases. For 16QAM signaling, the results show that FTN provides a concrete spectral-efficiency gain before the finite-alphabet saturation limit is reached. For example, under fixed $\mathsf{SNR}_{\rm rx}=20$ dB in Fig.~\ref{fig:captaurx}, reducing $\delta$ from $1$ to $0.1$ increases the rate from about $3.0$ to nearly $3.9$ bits/s/Hz, demonstrating a practical gain of DFT-precoded FTN with finite-alphabet inputs. The results also demonstrated that the capacity behavior depends strongly on the SNR definition: under fixed $\mathsf{SNR_{tx}}$, the rate saturates after $\delta$ decreases below a threshold, whereas under fixed $\mathsf{SNR_{rx}}$, decreasing $\delta$ increases the physical transmit power and can further improve the information rate. 
In this paper, we investigated the constrained capacity of DFT-precoded FTN signaling over FS fading channels with finite-alphabet inputs. By employing CP and CS, the combined FTN ISI and multipath channel was decomposed into parallel eigenchannels, enabling constrained-capacity analysis under practical constellations. We further considered the case without CP/CS and formulated the mismatched AIR caused by finite-block imperfect diagonalization and residual inter-eigenchannel interference. For 16QAM signaling, DFT-precoded FTN achieves clear spectral-efficiency gains over Nyquist before reaching the finite-alphabet saturation limit. Simulation results showed that the mismatched AIR suffers a loss at short block lengths but gradually approaches the constrained capacity as $N$ increases. Finally, the mismatch loss is more pronounced at low SNR, where noise and residual inter-eigenchannel interference have a stronger impact on the decoding metric. Future work includes extending the results to higher order QAM modulations. 
\vspace{-0.2cm} \enlargethispage{-0.38in}

\bibliographystyle{IEEEtran}

\bibliography{main}

\end{document}